\documentclass[aps,prl,twocolumn,superscriptaddress,floatfix,showpacs]{revtex4-1}

\usepackage{graphicx}
\usepackage{amsmath}

\begin{document}

\title{Energetic molding of chiral magnetic bubbles}

\author{Derek Lau}
\affiliation{Department of Materials Science \& Engineering, Carnegie Mellon University, Pittsburgh, Pennsylvania 15213, USA}

\author{Vignesh Sundar}
\author{Jian-Gang Zhu}
\affiliation{Department of Electrical \& Computer Engineering, Carnegie Mellon University, Pittsburgh, Pennsylvania 15213, USA}

\author{Vincent Sokalski}
\email[]{vsokalsk@andrew.cmu.edu}
\affiliation{Department of Materials Science \& Engineering, Carnegie Mellon University, Pittsburgh, Pennsylvania 15213, USA}

\date{\today}

\begin{abstract}

Topologically protected magnetic structures such as skyrmions and domain walls (DWs) have drawn a great deal of attention recently due to their thermal stability and potential for manipulation by spin current, which is the result of chiral magnetic configurations induced by the interfacial Dzyaloshinskii-Moriya Interaction (DMI). Designing devices that incorporate DMI necessitates a thorough understanding of how the interaction presents and can be measured.  One approach is to measure growth asymmetry of chiral bubble domains in perpendicularly magnetized thin films, which has been described elsewhere by thermally activated DW motion. Here, we demonstrate that the anisotropic angular dependence of DW energy originating from the DMI is critical to understanding this behavior. Domains in Co/Ni multi-layers are observed to preferentially grow into non-elliptical teardrop shapes, which vary with the magnitude of an applied in-plane field. We model the domain profile using energetic calculations of equilibrium shape via the Wulff construction, which explains both the teardrop shape and the reversal of growth symmetry at large fields. 

\end{abstract}

\pacs{75.70.Cn, 75.70.Kw, 65.40.gp,75.70.-i}

\maketitle

The ability to manipulate chiral magnetic structures via the spin Hall effect has sparked renewed interest in spintronic devices that may lead to lower power, non-volatile memory and logic circuits.\cite{Bromberg2014,Emori2013,Fukami2009,Parkin2008,Ryu2014} A key challenge is accurately measuring the Dzyaloshinskii-Moriya Interaction (DMI) and exploring interfaces that produce a larger net interfacial DMI in magnetic films to improve the efficiency by which novel spin configurations can be stabilized and manipulated with electric current. The DMI energy, $E=-\vec{D}\cdot (\vec{S}_1\times \vec{S}_2)$, originates from the spin-orbit interaction and is analogous to the better known Heisenberg exchange interaction, $E=-J(\vec{S}_1\cdot \vec{S}_2)$.\cite{Dzyaloshinsky1958,Moriya1960} While Heisenberg exchange favors parallel or anti-parallel arrangement of neighboring spins, $\vec{S}_1$ and $\vec{S}_2$, the DMI favors orthogonal alignment.  For a polycrystalline thin film with inversion symmetry broken by dissimilar seed and capping layers, the DMI vector simplifies to $\vec{D}=D_{int}(\hat{r}\times\hat{z})$ according to Moriya's rules\cite{Moriya1960,Thiaville2012}, where $\hat{r}$ is the unit vector connecting $\vec{S}_1$ to $\vec{S}_2$ and $\hat{z}$ is along the film normal. This interfacial DMI, $D_{int}$, explains the formation of skyrmions and novel spin textures in magnetic thin films that have canted spin structures.\cite{Bode2007,Heide2008,Heinze2011,Huang2012,Jiang2015,Yu2010} Another consequence of interfacial DMI is the preference for formation of N\'{e}el domain walls (DWs) with preferred chirality over magnetostatically favored Bloch walls. In this context, the interfacial DMI is treated as a field, $H_{DMI}$, acting perpendicular to the DW causing a rotation from Bloch to N\'{e}el type, where $\mu _{o}H_{DMI}=D_{int}/M_s\lambda$. \cite{Chen2013,Hrabec2014,Je2013} 

\begin{figure}
\includegraphics[width = 3.4in]{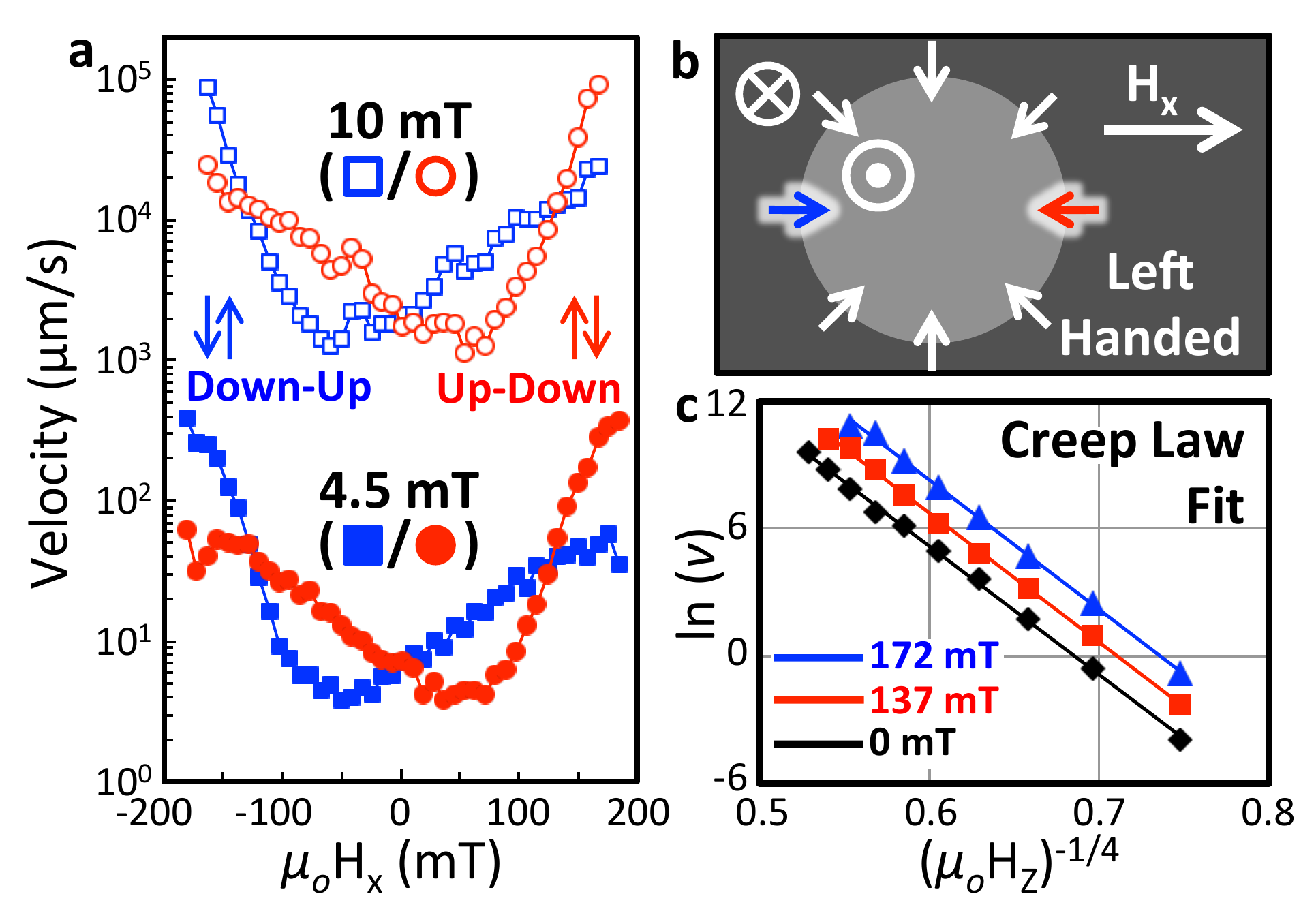}
\caption{\label{VelocityPlot3}a)Domain wall velocity as a function of $\mu _{o}H_x$ for $\mu _{o}H_z =$ 4.5mT (closed) and 10mT (open) using pulse widths of 100ms and 5ms, respectively. b)Bubble schematic with the inferred left-handed N\'{e}el wall chirality highlighting the magnetization vector within the Down-Up (blue) and Up-Down (red) DWs. c)Creep law dependence of growth velocity on $H_z$ for the Up-Down wall with varying $\mu _{o}H_x$ as indicated.}
\end{figure}

\begin{figure*}
\includegraphics[width = 7in]{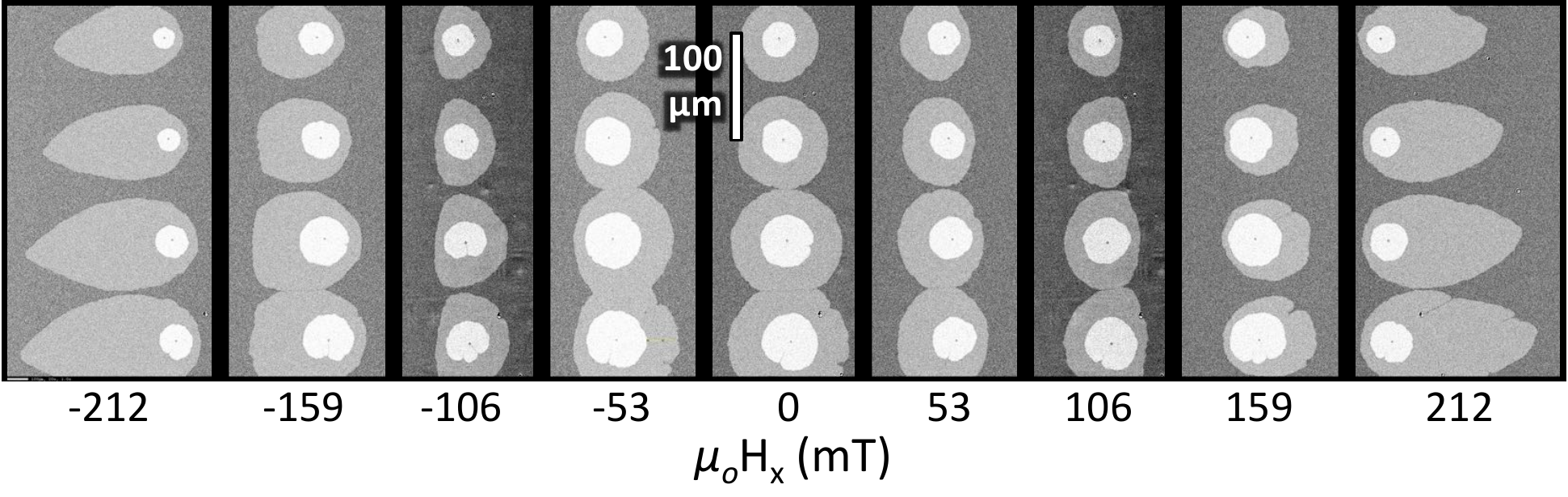}
\caption{\label{KerrImages}Superimposed perpendicular Kerr images of Co/Ni multi-layer films showing impact of $\mu _{o}H_x$ on domain morphology for an array of four nucleated bubbles separated by 100$\mu$m.  White corresponds to the initial perpendicular domain oriented out of the page and nucleated with $\mu _{o}H_x=$ 0.  Light grey is the domain shape that resulted after applying a 10mT perpendicular field with varying $\mu _{o}H_x$ as indicated.  Dark grey indicates magnetization oriented into the page.}
\end{figure*}

The growth asymmetry of perpendicular chiral magnetic domains in the presence of an externally applied planar magnetic field, $H_{x}$, was first suggested to originate from the DMI by Kabanov et al \cite{Kabanov2010} and later described in the context of creep DW motion using Co thin films grown on Pt.\cite{Hrabec2014,Je2013,Lavrijsen2015,Petit2015} The creep velocity is given by $V=V_o\exp [-\alpha ^{*}(\mu _oH_z)^{-1/4}]$ where $\alpha^*$ is directly related to the energy of the DW, which will be a minimum (maximum) when $H_{DMI}$ is parallel (anti-parallel) to the applied field, $H_x$. According to this model, the wall velocity, $v$ vs. $\mu _{o}H_x$, should be shifted for both the left and right sides of a magnetic bubble with a minimum occurring where $H_{x}$ cancels $H_{DMI}$.  However, it has frequently been observed that such a curve does not only shift, but also takes on an asymmetric shape.  This has been attributed to dependence of DW width on applied field direction\cite{Kim2016} although the magnitude of this effect is small.  Another explanation is chiral damping, which has also been speculated to exist due to the spin-orbit interaction.\cite{Jue2016}  While each of these factors could contribute to the asymmetric DW velocity, the role of the overall domain shape and angular dependence of wall energy have been largely ignored.  Here, we perform experimental studies on chiral magnetic bubbles in Co/Ni multi-layers prepared on a Pt seedlayer to understand the evolution of the domain shape during growth.  The unique shapes and anomalous growth symmetry observed experimentally are explained through equilibrium calculations based on minimization of domain wall energy using the Wulff construction.\cite{Herring1951,Wulff1901}

Films were prepared on Si(001) substrates with native oxide by DC magnetron sputtering from 5 inch targets in an Argon atmosphere fixed at 2.5mTorr unless otherwise noted.  Base pressure was maintained at $< 3\times10^{-7}$ Torr.  The film stack examined in this work is Si/TaN(3)/Pt(2.5)/[Co(0.2)/Ni(0.6)]$_2$ /Co(0.2)/Ta(0.5)/TaN(3) with units in nm.  The TaN layers were prepared by reactive sputtering in an Ar(2.5mTorr)/N$_2$(0.5mTorr) gas mixture.  M-H loops measured by alternating gradient field magnetometry(AGFM) and vibrating sample magnetometry (VSM) indicate $M_s$ = 600 kA/m, perpendicular $\mu_oH_c = 10 $mT, and in-plane $\mu_oH_k = 1$T.  From this, $K_{eff}$ is calculated as $\mu_oH_kM_s/2 = 3\times10^5 J/$m$^3$.  Films were confirmed to have FCC(111) fibre-texture by x-ray diffraction (XRD) as required for perpendicular magnetic anisotropy in most Co-based multi-layers.\cite{Kurt2010}  The nucleation and growth of magnetic domains were observed using a white light, wide-field Kerr microscope.  An electromagnet was used to apply in-plane fields while a perpendicular coil supplied short field pulses.  Prior to observation, a focused Ga$^+$ beam was used to damage a spot on the film with a dose of 1800 $\mu C/$cm$^2$ (at 5kV accelerating voltage) creating a nucleation site for magnetic bubble domains.  Images were analyzed digitally to determine domain dimensions and normalized by the pulse length to determine average growth velocity.  The dependence of growth velocity on in-plane field was performed by initially nucleating a magnetic domain of 10$\mu$m diameter followed by perpendicular field pulses as indicated.  For slow moving domain walls, multiple pulses were used to produce an appreciable displacement and it was observed that each subsequent pulse produced the same displacement.  

Figure \ref{VelocityPlot3} shows the domain wall velocity for the left (Down-Up) and right (Up-Down) side of a magnetic bubble as a function of applied field, $H_x$, for $\mu_oH_z=$ 4.5 and 10mT. Also shown is the linear trends in ln(v) vs $(\mu _{o}H_z)^{-1/4}$ confirming creep behavior for all scenarios studied.  While the velocity increased by more than two orders of magnitude for $\mu _{o}H_z=$ 10mT, the overall trend remains similar.  A minimum in velocity occurs at positive $\mu _{o}H_{x}$ for the Up-Down domain wall implying left-handed chirality (Figure \ref{VelocityPlot3}b) according to the creep law, which is consistent with the sign of DMI measured and calculated elsewhere for Co/Pt interfaces.\cite{Hrabec2014,Je2013,Lavrijsen2015,Vanatka2015,Yang2015} We also note that the curves are highly asymmetric about their minima resulting in a reversal of the preferred growth direction beyond some $H_x$.  If the chirality predicted according to the creep law at low fields is correct, this suggests that at large fields, the domain wall where $H_{DMI}$ is anti-parallel to $H_x$ displays a greater velocity; something that has not previously been reported.  Moreover, the shape of the magnetic domain evolves with increasing $H_x$ as shown in Figure \ref{KerrImages} for an array of four bubbles studied simultaneously.  Most notably, for $H_x$ beyond the value at which the anomalous growth behavior is observed, the domain adopts a teardrop shape.  We also note the formation of flattened shapes at lower $H_x$.

\begin{figure}
\includegraphics[width = 3.4in]{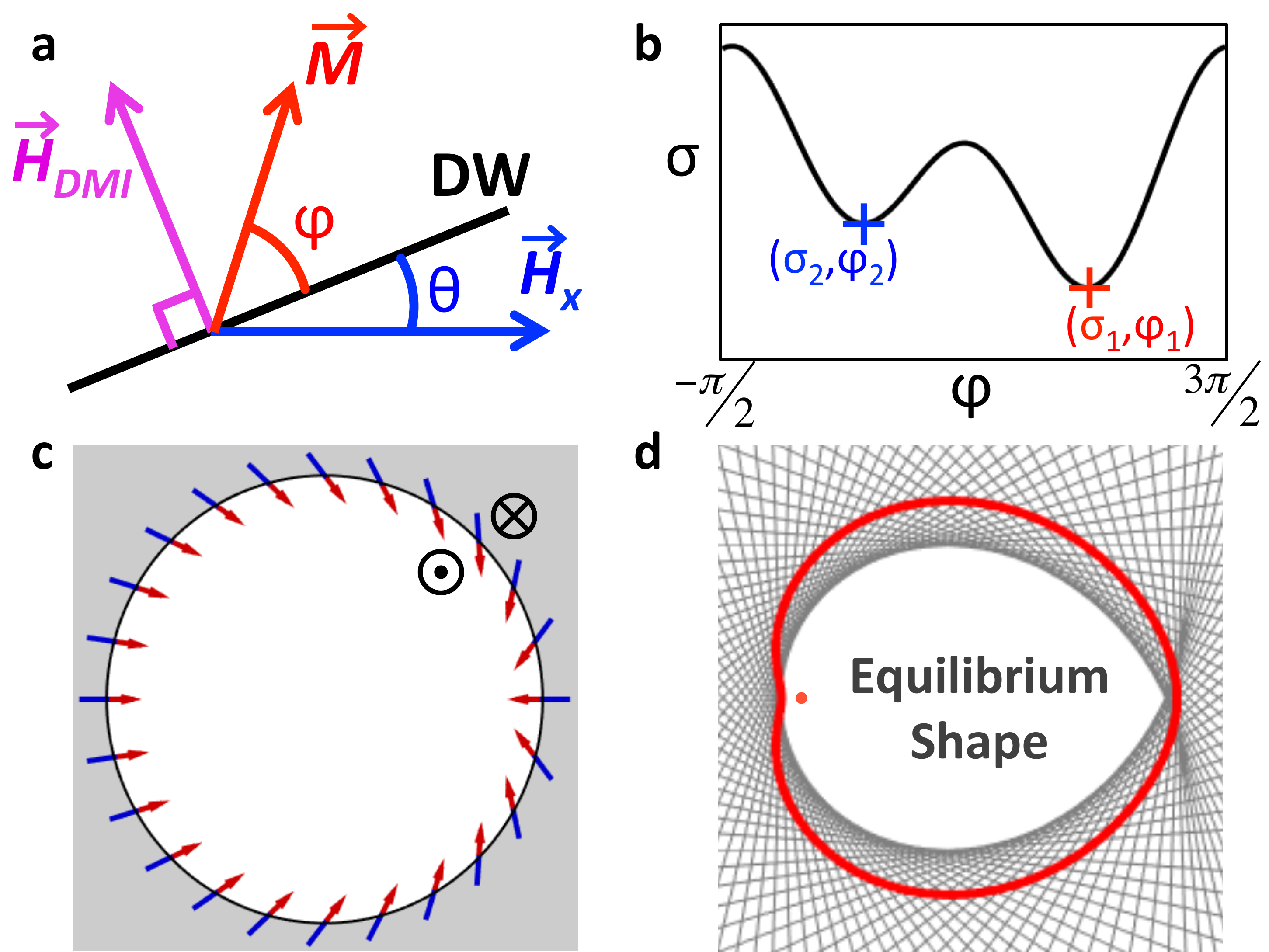}
\caption{\label{ModelSchematic}Calculation methodology using Wulff Construction. a) Coordinate system. b) Example DW energy for fixed $\theta$ showing possibility of two energy minima. c) Example magnetic moment configuration within DW for $\mu _{o}H_{x}$ = 240mT, $\mu _{o}H_{DMI}$ = 300mT, $K_{eff} = 3\times10^5 $J/m$^3, A = 1\times10^{-11} $J/m$, M_s = 6 \times10^5 $A/m, and $K_D = 3\times10^4 $J/m$^3$.  d) $\sigma(\theta)$ polar energy plot (red) and Wulff Construction (grey) resulting in equilibrium shape corresponding to (c). The origin of the polar energy plot is indicated by the red point.}
\end{figure}

We now shift gears to better understand the origin of the anamolous growth symmetry in Figure \ref{VelocityPlot3} and the notable evolution of the domain shape as a function of applied field.  The presence of cusps in the teardrop domains of Figure \ref{KerrImages} show a striking resemblance to the facets and edges seen in bulk single crystals, which exist as a consequence of highly anisotropic interface energy.  To address this in chiral bubbles, we consider the total free energy of the magnetic domain wall as a function of its orientation with respect to $H_x$.  The expression for DW energy is given by the following:\cite{Je2013,Pizzini2014} 
\begin{multline}
  \sigma(\varphi ,\theta)=\sigma _{o}-\pi\lambda \mu_oH_xM_s\cos (\varphi+\theta)- 
 \\ 
  \pi\lambda \mu_oH_{DMI}M_s\sin (\varphi)+2\lambda K_D\sin^2(\varphi)$$ 
\end{multline}

Where $\theta$ and $\varphi$ are defined in Figure \ref{ModelSchematic}a. $\sigma_o=4\sqrt{AK_{eff}}$ is the Bloch wall energy and $\lambda=\sqrt{A/K_{eff}}$ is the domain wall width.  $K_D$ is the domain wall anisotropy energy that originates from the magnetostatic favorability to form Bloch walls in the absence of DMI.  For a given value of $\theta$, the equilibrium direction of magnetization $\varphi$ is determined along with the corresponding energy, $\sigma$.  Therefore, it is possible to produce an equilibrium polar energy plot $\sigma(\theta)$ as shown in Figure \ref{ModelSchematic}d (red), which is used to calculate the equilibrium shape via the Wulff construction.  

\begin{figure}
\includegraphics[width = 3.4in]{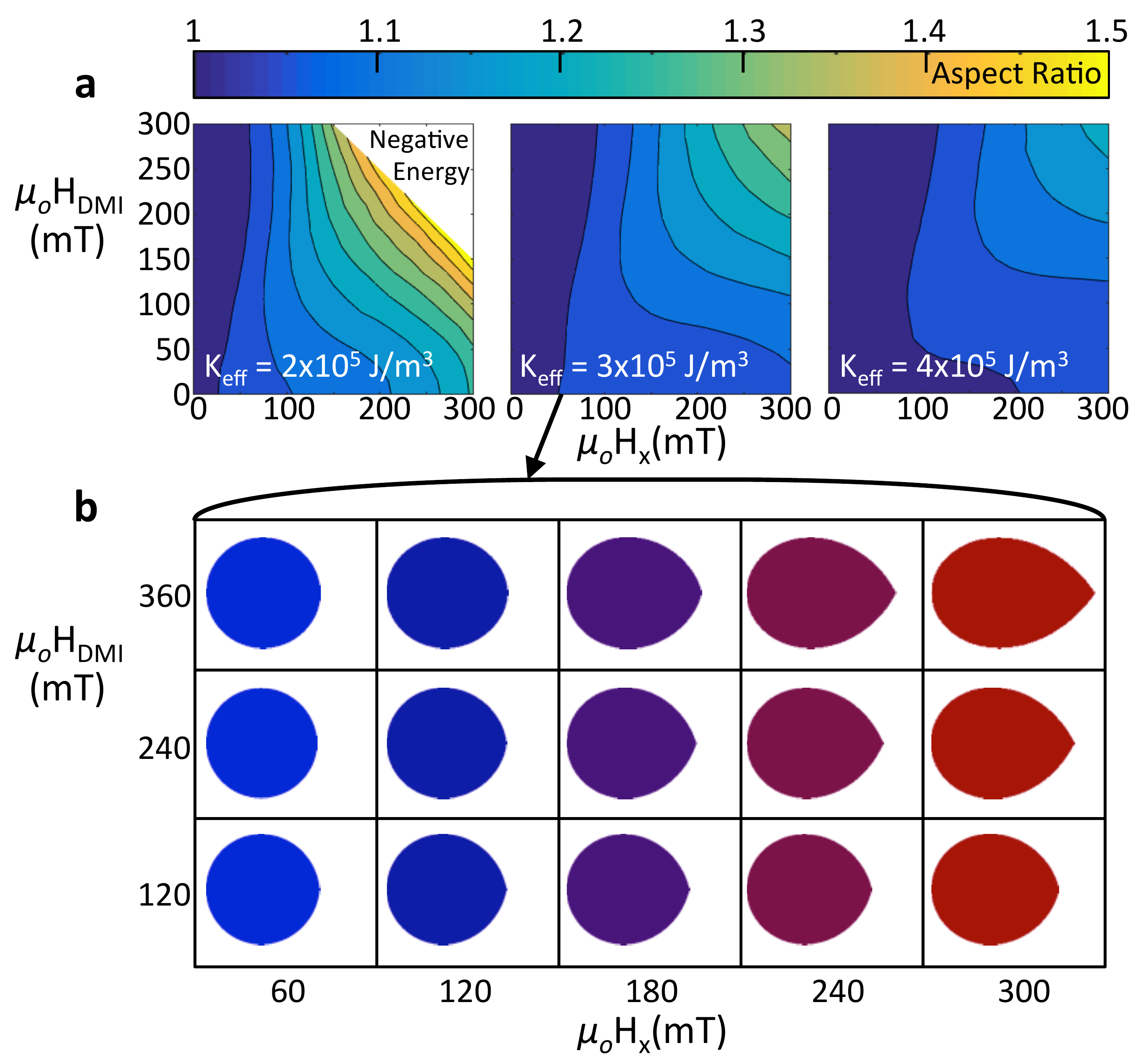}
\caption{\label{EquilibriumResult}a) Contour plots of magnetic domain aspect ratio calculated using the Wulff construction as a function of $K_{eff}$, $H_{DMI}$, and $H_x$.  The effect is most pronounced for small $K_{eff}$.  b) Domain shape calculated using the equilibrium Wulff construction as a function of $H_{DMI}$ and $H_x$. $K_{eff} = 3\times10^5 J/m^3, A = 1\times10^{-11} J/m, M_s = 6 \times10^5 A/m, K_D = 3\times10^4 J/m^3$.}
\end{figure}

The impact of $H_x$ and $H_{DMI}$ on equilibrium shape determined from the Wulff construction are outlined in Figure \ref{EquilibriumResult} using experimentally measured values of $K_{eff}$ and $M_s$. Here, we have assumed exchange stiffness, $A = 1\times10^{-11} J/m$ and domain wall anisotropy constant, $K_D = 3\times10^4 J/m^3$.\cite{Yoshimura2016}  However, for equilibrium shape calculations, it can be shown that the resulting shape is independent of $A$ and weakly dependent on $K_D$ for realistic values.  It's clear that as $H_x$ approaches $H_{DMI}$, the equilibrium shape is pulled along one direction forming a high angle cusp as part of the teardrop shape.\cite{Boojum}  This can be explained by the domain attempting to minimize the density of those walls that have $H_{DMI}$  anti-parallel to $H_x$.  While the Wulff construction does not calculate DW velocity directly, it does identify an anisotropic driving force for growth that is favored along the direction where the teardrop cusp is located. Moreover, we note that $H_{DMI}$ predicted from the velocity minimum in figure 1 of 60mT is significantly smaller than the values necessary to stabilize the teardrop shape based on our calculation.

\begin{figure}
\includegraphics[width = 3.2in]{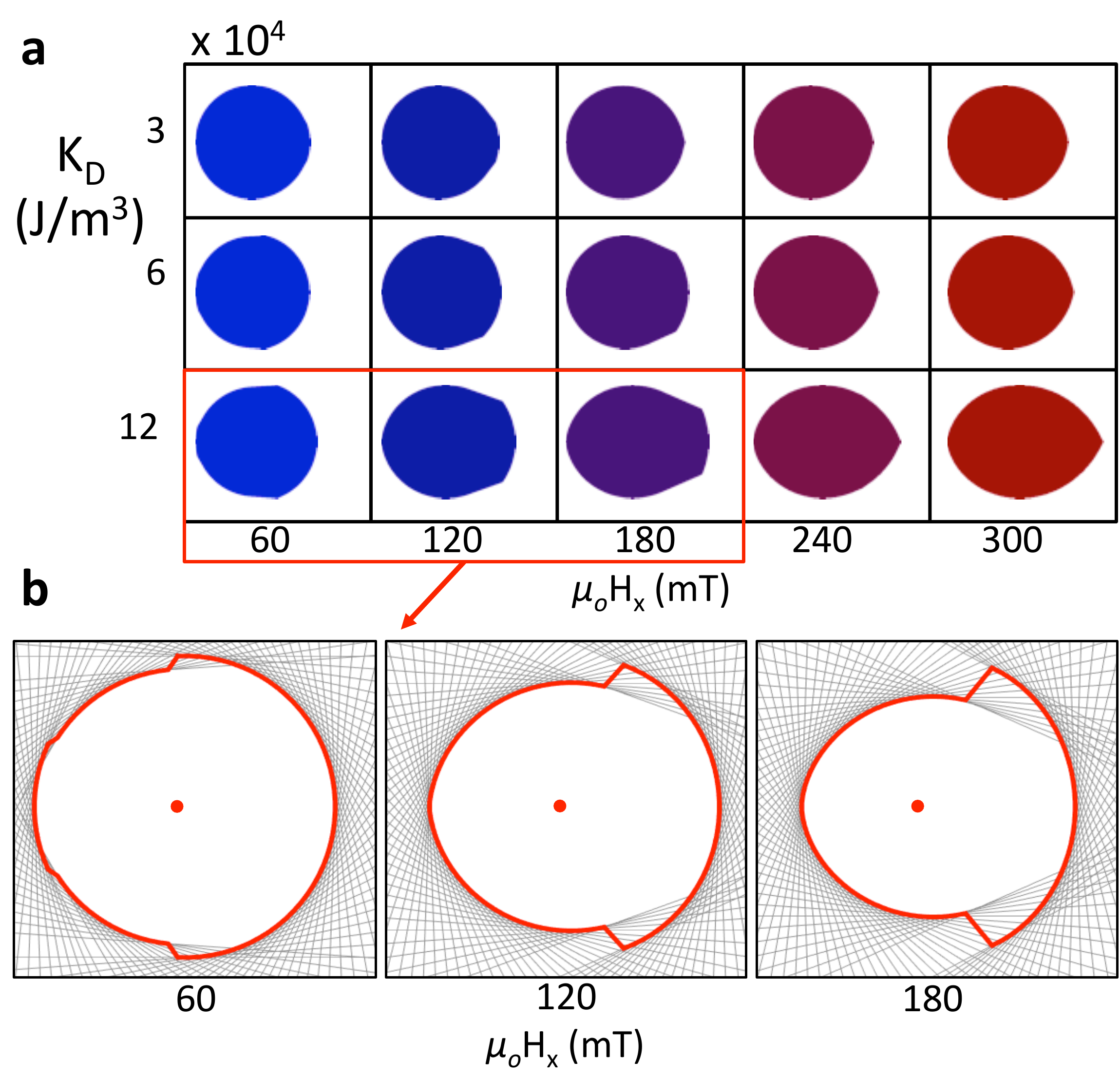}
\caption{\label{PseudoEquilibrium}a) Domain shape calculated from the pseudo-equilibrium Wulff construction as a function of $H_x$ for varied domain wall anisotropy energy, $K_D$.  $\mu_oH_{DMI} = 120 mT, K_{eff} = 3\times10^5 J/m^3, M_s = 600 kA/m, A = 1\times10^{-11} J/m$.  b) Combined polar energy plots with resulting Wulff construction using the pseudo-equilibrium approximation for $K_D = 12\times10^4 J/m^3$.  Red points represent the center of the polar energy plot.  The sharp change in energy is associated with the stabilization of a metastable orientation variant.}
\end{figure}

The teardrop shape has also been observed elsewhere in magnetic thin films\cite{Hrabec2014,Lavrijsen2015,Woo2016}, but no evidence has been presented to explain its origin or that suggests it coincides with a reversal in growth symmetry.  We note, however, that the material parameters associated with Co/Ni multi-layers in our case deviate significantly from the Pt/Co bi-layers studied in these works. Specifically, the smaller $K_{eff}$ in our films will decrease the resting Bloch wall energy $\sigma_o$ while increasing the impact of Zeeman energy terms from $H_{DMI}$ and $H_x$ via increasing $\lambda$.  One additional consequence of small $K_{eff}$  is that negative wall energies are realized for the field values under consideration here.  This suggests an instability of uniform perpendicular magnetization, which has been predicted to cause relaxation towards chiral spin textures or skyrmion lattices.\cite{Bogdanov2005,Rohart2013,Pizzini2014}  Such negative wall energies produce a wind in the polar free energy plot, which greatly complicates interpretation of the Wulff construction. The impact of $K_{eff}$ on equilibrium shape is shown in the contour plots of Figure \ref{EquilibriumResult}.  Here, the aspect ratio, which originates primarily from the teardrop elongation, is plotted as a function of $H_x$ and $H_{DMI}$.  In all cases, this aspect ratio increases with increasing $H_x$  or $H_{DMI}$, particularly when the two fields are comparable in magnitude.  Most notably, as $K_{eff}$ is reduced the aspect ratio increases for any combination of $H_x$ and $H_{DMI}$.  This could explain why the anomalous growth direction and teardrop shapes observed in this work have not been reported in comparable cases where $K_{eff}$ is larger.

While the Wulff construction explains the formation of teardrop shapes at high field, it does not explain the flattened  shape at low $H_x$.  In fact, it is counter to such a result as this shape increases the density of domain walls that have $H_{DMI}$ anti-parallel to the applied field.  Instead, we demonstrate how this shape could be explained through a refinement of our calculation we call a pseudo-equilibrium Wulff construction.  From Figure \ref{ModelSchematic}b, it is clear that for some values of $\theta$, there can be two energy minima associated with appreciable DW anisotropy energy.  In this approximation, we assume that the appearance of each variant will occur with equal probability.  As such, we have repeated our Wulff construction so that the energy of a domain wall is given by the average energy of the two possible magnetization directions.  If there is only a single energy minimum, then that value represents the equilibrium DW energy.  The results of this construction are shown in Figure \ref{PseudoEquilibrium}.  It can be seen that for large $K_D$ and low $H_x$, the flattening observed in experiment is reproduced.  This occurs because $K_D$ stabilizes the DWs with multiple variants, which will have higher energy than those that do not.  Therefore, the equilibrium bubble shape maximizes length for DWs that do not form metastable variants.  For example, consider Figure \ref{PseudoEquilibrium}b where the left arc is free of multiple variants while the right is not.  As $H_x$ increases, the fraction of DWs that can support multiple variants shrinks along with the length of the flattened region.  Beyond some value of $H_x$, multiple variants are no longer supported or only supported for a narrow region on the right side of the bubble and the equilibrium construction is recovered.

However, the flattening is only observed for unrealistically large values of $K_D$ and, therefore, the situation requires a more complicated explanation.  The formation of Bloch points in perpendicular Co/Ni multi-layers have been discussed in much detail elsewhere and are predicted to form in significant density where $H_{DMI}$ is anti-parallel to $H_x$.  Such Bloch points would increase the wall energy where stable and also serve as pinning sites during DW growth.\cite{Yoshimura2016}  It is possible that the rapid transition from a flat growth front to a teardrop shape is related to the annihilation of Bloch points at a critical field that had been pinning the DW.  Such a critical field would be directly related to the interfacial DMI vector although additional micromagnetic modeling is needed to quantitatively predict Bloch point density and determine the contribution to overall DW energy and growth dynamics. 

In summary, we have presented a new paradigm for describing the behavior of chiral magnetic bubbles that is based on classical interface thermodynamics, which has been paramount in the broader field of phase transformations, but has not previously been applied in the context of chiral magnetism.  We show that the non-elliptical teardrop morphology associated with growth of magnetic domains in structurally asymmetric Co/Ni thin films can be explained by a driving force towards the formation of equilibrium shapes. Furthermore, the appearance of teardrop domains coincides with a reversal of the growth symmetry expected from creep motion dynamics. Both the teardrop shape and the anomalous preferred growth direction is predicted from the Wulff construction and becomes most pronounced as $H_x$  approaches $H_{DMI}$.  Our study here demonstrates that the wall velocity obtained by static wall position measurements after motion is actually the combined effect of thermally activated creep driven by perpendicular magnetic fields and the energetic relaxation of the bubble domain towards its equilibrium shape during growth.  This new driving force suggests that $H_{DMI}$ determined from the minimum in velocity underestimates the magnitude of the DMI vector, especially for low $K_{eff}$ materials where the polar energy plots become highly anisotropic.  This may explain discrepancies in growth asymmetry observed throughout the literature and inconsistencies between direct observations of wall chirality and those inferred from asymmetric bubble growth. 

This work was partially supported by the Collaborative Innovation Research Center through The Joint Institute of Engineering at Sun Yat-Sen University, The Samsung Global MRAM Innovation Project, and the Berkman Faculty Development Fund.

\bibliography{PRL_Citations_Rep}

\end{document}